\documentclass[10pt,msam,msbm]{iopart}
\usepackage{iopams}
\usepackage{graphicx}
\usepackage{dcolumn}
\usepackage{bm}
\usepackage{color}

\begin{document}

\title[]{3D-xy critical properties of YBa$_{2}$Cu$_{4}$O$_{8}$ and magnetic field induced 3D to 1D crossover}

\author{S~Weyeneth$^1$, T~Schneider$^1$, Z~Bukowski$^2$, J~Karpinski$^2$ and H~Keller$^1$}

\address{$^1$~Physik-Institut der Universit\"{a}t Z\"{u}rich, Winterthurerstrasse 190, CH-8057 Z\"urich, Switzerland}
\address{$^2$~Laboratory for Solid State Physics, ETH Z\"urich, CH-8093 Z\"{u}rich, Switzerland}

\ead{wstephen@physik.uzh.ch}
\begin{abstract}
We present reversible magnetization data of a YBa$_{2}$Cu$_{4}$O$_{8}$ single crystal and analyze the evidence for 3D-xy critical behavior and a magnetic field induced 3D to 1D crossover. Remarkable consistency with these phenomena is observed in agreement with a magnetic field induced finite size effect, whereupon the correlation length transverse to the applied magnetic field cannot grow beyond the limiting magnetic length scale $L_{H}=\left( \Phi _{0}/\left( aH\right) \right) ^{1/2}$. By applying the appropriate scaling form we obtain the zero-field critical temperature, the 3D to 1D crossover, the vortex melting line and the universal ratios of the related scaling variables. Accordingly there is no continuous phase transition in the ($H,T$)-plane along the $H_{c2}$-lines as predicted by the mean-field treatment.

\end{abstract}

\pacs{74.25.Bt, 74.25.Ha, 74.40.+k}
\maketitle

\section{\label{sec:level1}Introduction}
Fluctuation effects are known to be strongly enhanced in the high temperature cuprate superconductors due to their anisotropic behavior and their high zero-field transition temperature $T_c$ \cite{book,larkin}. For YBa$_{2}$Cu$_{4}$O$_{8}$ and related compounds several studies pointed to the importance of critical fluctuations \cite{Lascialfari, Barbaduc, Rosenstein, Rigamonti}. To circumvent the smearing of the phase transition due to inhomogeneities YBa$_{2}$Cu$_{4}$O$_{8}$ is an exquisite candidate due to its nearly stoichiometric structure and the availability of excellent single crystals \cite{karpinsky,karp1999}. As YBa$_{2}$Cu$_{4}$O$_{8}$ is intrinsically doped there is no anomalous precursor diamagnetism expected, as reported for example in Aluminium doped MgB$_2$ \cite{Rigamonti}.\\
In this study we present reversible magnetization data of a YBa$_{2}$Cu$_{4}$O$_{8}$ single crystal and analyze the evidence for 3D-xy critical behavior and a magnetic field induced 3D to 1D crossover. We observe
remarkable consistency with these phenomena. Since near $T_c$ thermal fluctuations are expected to dominate \cite{book,jhts,parks,ts07,tsepl}, Gaussian fluctuations point to a magnetic
field induced 3D to 1D crossover \cite{lee}. Whereby the effect of
fluctuations is enhanced, it appears inevitable to take thermal fluctuations
into account. Indeed, invoking the scaling theory of critical phenomena we
show that the data are inconsistent with the traditional mean-field
interpretation. On the contrary, we observe agreement with a magnetic field
induced finite size effect, whereupon the correlation length transverse to
the magnetic field $H_{i}$, applied along the $i$-axis, cannot grow beyond
the limiting magnetic length
\begin{equation}
L_{H_{i}}=\left( \Phi _{0}/\left( aH_{i}\right) \right) ^{1/2},  \label{eq1}
\end{equation}
with $a\simeq 3.12$ \cite{bled}. $L_{H_{i}}$ is related to the average
distance between vortex lines. Indeed, as the magnetic field increases, the
density of vortex lines becomes greater. But this cannot continue
indefinitely. The limit is roughly set on the proximity of vortex lines by
the overlapping of their cores. This finite size effect implies that in type
II superconductors, superconductivity in a magnetic field is confined to
cylinders with diameter $L_{H_{i}}$\cite{tsepl,Weyeneth}. Accordingly, below $T_{c}$ there is the
3D to 1D crossover line
\begin{equation}
H_{pi}\left( T\right) =\left( \Phi _{0}/\left( a\xi _{j0}^{-}\xi
_{k0}^{-}\right) \right) (1-T/T_{c})^{4/3},  \label{eq2}
\end{equation}
with $i\neq j\neq k$. $\xi _{i0,j0,k0}^{- }$ denotes the critical
amplitudes of the correlation lengths below $T_{c}$ along the
respective axis. It circumvents the occurrence of the continuous phase
transition in the $(H_c,T)$-plane along the $H_{c2}$-lines predicted by the
mean-field treatment \cite{abrikosov}. Indeed, the relevance of thermal
fluctuations emerges already from the reversible magnetization data shown in
Fig. \ref{fig1}. As a matter of fact, the typical mean-field behavior \cite{abrikosov}, whereby the magnetization scales below $T_{c}$ linearly with
the magnetic field, does not emerge.\\

\section{\label{sec:level1}Experiment and Analysis}

\begin{figure}[t!]
\centering
\vspace{-0.6cm}
\includegraphics[width=0.7\linewidth]{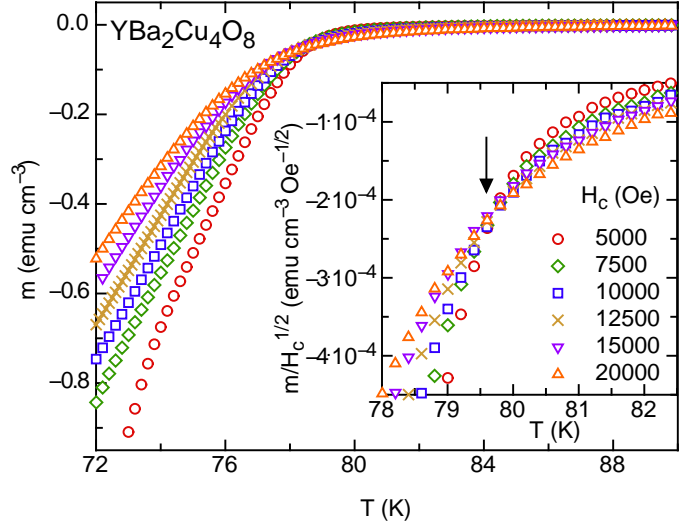}
\vspace{0cm} \caption{Reversible magnetization $m$ \textit{vs}. $T$ of a YBa$_2$Cu$_4$O$_8$ single crystal for magnetic fields $H_{c}$ applied along the $c$-axis. The inset shows $m/H_c^{1/2}$ \textit{vs}. $T$. The arrow indicates the crossing point which yields the estimate $T_c \simeq 79.6 K$.}
\label{fig1}
\end{figure}

\begin{figure}[t!]
\centering
\includegraphics[width=0.7\linewidth]{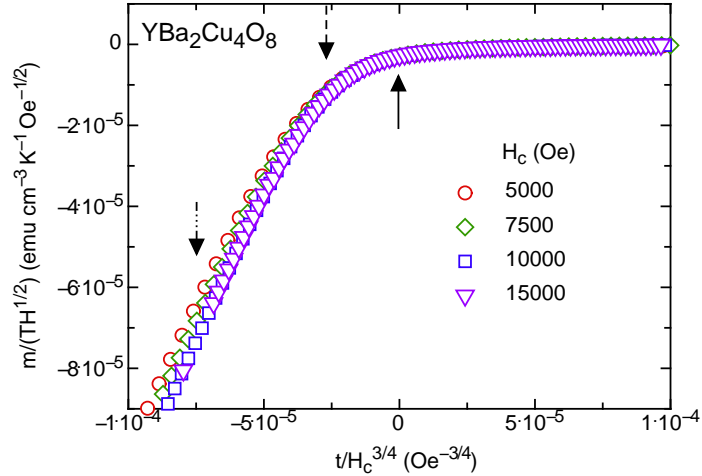} \vspace{0cm} \caption{$m/\left( TH_{c}^{1/2}\right) $ \textit{vs}. $t/H^{3/4}$ for a YBa$_2$Cu$_4$O$_8$ single crystal with $T_{c}=79.6$ K. The full arrow marks the zero-field critical temperature $T_c$, the dashed arrow the 3D to 1D crossover and the dotted arrow the vortex melting line.}
\label{fig2}
\end{figure}

\begin{figure}[t!]
\centering
\vspace{-0.6cm}
\includegraphics[width=0.7\linewidth]{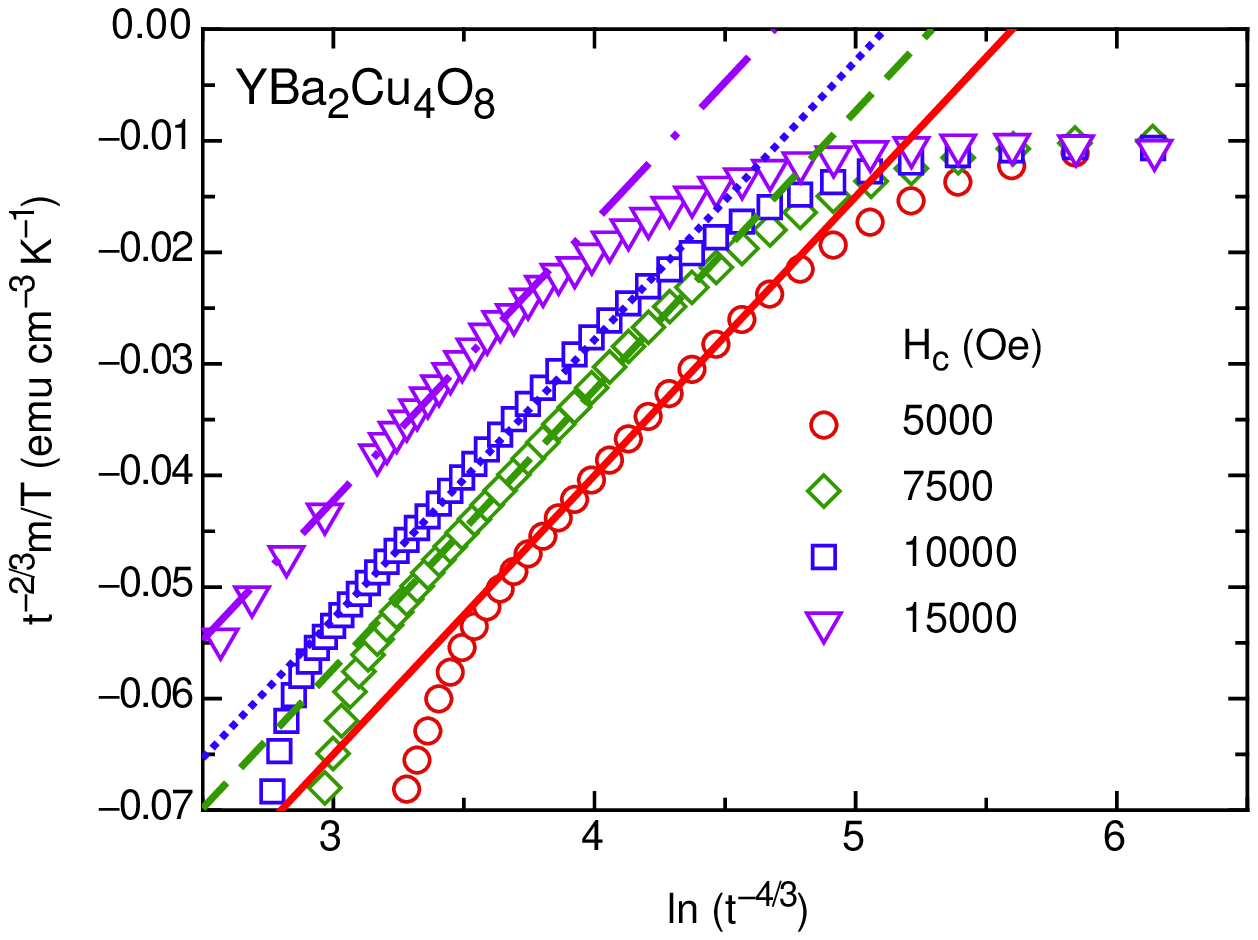} \vspace{0cm} \caption{$\left\vert t\right\vert ^{-2/3}m/T$ \textit{vs}. ln$\left(
\left\vert t\right\vert ^{-4/3}\right) $ for a YBa$_2$Cu$_4$O$_8$ single crystal according to Eq. (\ref{eq13}). The lines are fits to the rescaled magnetization data. Here $\xi _{ab0}^{2}\left\vert t_{p}\right\vert^{-4/3}=\Phi _{0}/aH_{c}$.}
\label{fig3}
\end{figure}

\begin{figure}[t!]
\centering
\vspace{-0.6cm}
\includegraphics[width=0.7\linewidth]{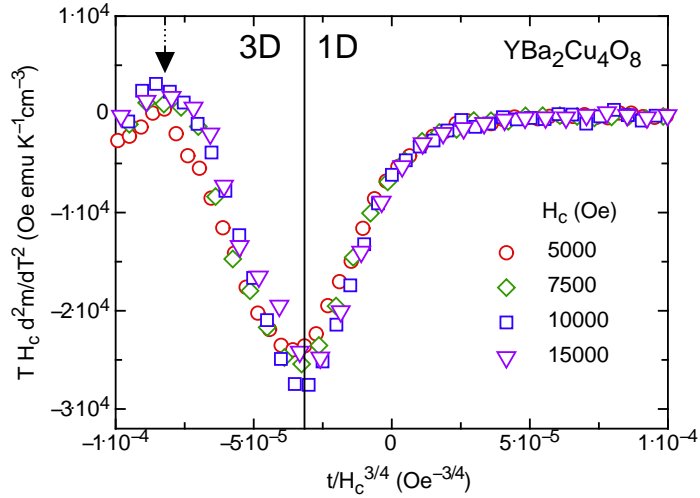}
\vspace{0cm} \caption{$TH_{c}d^{2}m/dT^{2}$ \textit{vs}. $x=t/H_{c}^{3/4}$ for a YBa$_{2}$Cu$_{4}$O$_{8}$ single crystal. The arrow marks the vortex melting line $x_{m}\simeq -8.35\cdot 10^{-5}($ Oe$^{-3/4})$ and the
vertical line $x_{p}\simeq -2.85\cdot 10^{-5}($ Oe$^{-3/4})$ the 3D to 1D
crossover line}
\label{fig4}
\end{figure}

The YBa$_{2}$Cu$_{4}$O$_{8}$ single crystal investigated in this work was
fabricated by a high-pressure synthesis method described in detail elsewhere
\cite{karpinsky, karp1999}. The volume of the nearly rectangular shaped sample is
estimated to be $3.9\cdot 10^{-4}$ cm$^{3}$. The magnetic moment was measured
by a commercial Quantum Design DC-SQUID magnetometer MPMS XL with installed
RSO option, allowing to achieve a resolution of $10^{-8}$ emu. For this
experiment the applied magnetic field was oriented along the $c$-axis of the
crystal. For different magnitudes of the field, temperature dependent
magnetization curves where measured. The zero-field cooled (ZFC) and field
cooled (FC) data have been compared in order to probe the reversible
magnetization only. The superconducting susceptibility was finally obtained
by correcting the measured data for the normal state and sample holder
contributions. In Fig. \ref{fig1} we depicted some of the measured magnetization curves $m$
\textit{vs}. $T$ for magnetic fields $H_{c}$ applied along the $c$-axis. On
a first glance the data fall on rather smooth curves, revealing that the
extraction of critical and crossover behavior requires a rather detailed
analysis.
When thermal fluctuations dominate and the coupling to the charge is
negligible the magnetization per unit volume, $m=M/V$, adopts the scaling
form\cite{book,jhts,parks,ts07}
\begin{eqnarray}
\frac{m}{TH^{1/2}} &=&-\frac{Q^{\pm }k_{B}\xi _{ab}}{\Phi _{0}^{3/2}\xi _{c}}%
F^{\pm }\left( z\right) ,\textrm{ }F^{\pm }\left( z\right) =z^{-1/2}\frac{%
dG^{\pm }}{dz},  \nonumber \\
z =x^{-1/2\nu} &=& \frac{\left( \xi _{ab0}^{\pm }\right) ^{2}\left\vert
t\right\vert ^{-2\nu }H_{c}}{\Phi _{0}}.  \label{eq3}
\end{eqnarray}
$Q^{\pm }$ is a universal constant and $G^{\pm }\left( z\right) $ a
universal scaling function of its argument, with $G^{\pm }\left( z=0\right)
=1$. $\gamma =\xi _{ab}/\xi _{c}$ denotes the anisotropy, $\xi _{ab}$ the
zero-field in-plane correlation length and $H_{c}$ the magnetic field
applied along the $c$-axis. In terms of the variable $x$ the scaling form (\ref{eq3}) is similar to Prange's \cite{prange} result for Gaussian fluctuations. Approaching $T_{c}$ the in-plane correlation length diverges
as
\begin{equation}
\xi _{ab}=\xi _{ab0}^{\pm }\left\vert t\right\vert ^{-\nu },\textrm{ }t=T/T_{c}-1,\textrm{ }\pm =sgn(t).  \label{eq4}
\end{equation}
Supposing that 3D-xy fluctuations dominate the critical exponents are given
by \cite{pelissetto}
\begin{equation}
\nu \simeq 0.671\simeq 2/3,\textrm{ }\alpha =2\nu -3\simeq -0.013,  \label{eq5}
\end{equation}
and there are the universal critical amplitude relations \cite%
{book,jhts,parks,ts07,tsepl,pelissetto}
\begin{equation}
\frac{\xi _{ab0}^{-}}{\xi _{ab0}^{+}}=\frac{\xi _{c0}^{-}}{\xi _{c0}^{+}}%
\simeq 2.21,\textrm{ }\frac{Q^{-}}{Q^{+}}\simeq 11.5,\textrm{ }\frac{A^{+}}{A^{-}}=1.07,  \label{eq6}
\end{equation}
and
\begin{eqnarray}
A^{-}\xi _{a0}^{-}\xi _{b0}^{-}\xi _{c0}^{-} &\simeq &A^{-}\left( \xi
_{ab0}^{-}\right) ^{2}\xi _{c0}^{-}=\frac{A^{-}\left( \xi _{ab0}^{-}\right)
^{3}}{\gamma }  \nonumber \\
&=&\left( R^{-}\right) ^{3},\textrm{ }R^{-}\simeq 0.815.  \label{eq7}
\end{eqnarray}
$A^{\pm }$ is the critical amplitude of the specific heat singularity,
defined as
\begin{equation}
c=\frac{C}{Vk_{B}}=\frac{A^{\pm }}{\alpha }\left\vert t\right\vert ^{-\alpha
}+B,  \label{eq8}
\end{equation}
where $B$ denotes the background. Furthermore, in the 3D-xy universality
class $T_{c}$, $\xi _{c0}^{-}$ and the critical amplitude of the in-plane magnetic field
penetration depth $\lambda _{ab0}$ are not independent, but related by the
universal relation \cite{book,jhts,parks,ts07,pelissetto},
\begin{equation}
k_{B}T_{c}=\frac{\Phi _{0}^{2}}{16\pi ^{3}}\frac{\xi _{c0}^{-}}{\lambda
_{ab0}^{2}}=\frac{\Phi _{0}^{2}}{16\pi ^{3}}\frac{\xi _{ab0}^{-}}{\gamma
\lambda _{ab0}^{2}}.  \label{eq9}
\end{equation}
Furthermore, the existence of the magnetization at $T_{c}$, of the
penetration depth below $T_{c}$ and of the magnetic susceptibility above $
T_{c}$ imply the following asymptotic forms of the scaling function \cite{book,jhts,parks,ts07,tsepl}
\begin{eqnarray}
Q^{\pm }\left. \frac{1}{\sqrt{z}}\frac{dG^{\pm }}{dz}\right\vert
_{z\rightarrow \infty } &=&Q^{\pm }c_{\infty }^{\pm },  \nonumber \\
Q^{-}\left. \frac{dG^{-}}{dz}\right\vert _{z\rightarrow 0}
&=&Q^{-}c_{0}^{-}\left( \ln z+c_{1}\right) ,  \nonumber \\
Q^{+}\left. \frac{1}{z}\frac{dG^{+}}{dz}\right\vert _{z\rightarrow 0}
&=&Q^{+}c_{0}^{+},  \label{eq10}
\end{eqnarray}
with the universal coefficients
\begin{equation}
Q^{-}c_{0}^{-}\simeq -0.7,\textrm{ }Q^{+}c_{0}^{+}\simeq 0.9,\textrm{ }Q^{\pm
}c_{\infty }^{\pm }\simeq 0.5, \nonumber
\end{equation}
\begin{equation}
c_{1}\simeq 1.76.  \label{eq11}\vspace{-0.75cm}
\end{equation}
We are now prepared to analyze the magnetization data. To estimate $T_{c}$
we note that according to Eqs. (\ref{eq3}), (\ref{eq10}) and (\ref{eq11})
the plot $m/H_{c}^{1/2}$ \textit{vs}. $T$ should exhibit a crossing point at
$T_{c}$ because $m/TH_{c}^{1/2}$ tends to the value $%
m/T_{c}H_{c}^{1/2}=-0.5k_{B}\gamma \Phi _{0}^{-3/2}$. The inset in Fig. \ref{fig1}
reveals that there is a crossing point near $T_{c}\simeq 79.6$ K. Given this
estimate consistency with 3D-xy critical behavior then requires according to
the scaling form (\ref{eq3}) that the data plotted as $m/(TH_{c}^{1/2})$ \textit{vs}. $tH_{c}^{-1/2\nu }\simeq tH_{c}^{-3/4}$ should collapse near $tH_{c}^{-3/4}\rightarrow 0$ on a single curve. Evidence for this collapse
emerges from Fig. \ref{fig2} with $T_{c}\simeq 79.6$ K.
Considering the limit $z\rightarrow 0$ below $T_{c}$ the appropriate scaling
form is
\begin{equation}
\frac{m}{T}=-\frac{Q^{-}c_{0}^{-}k_{B}}{\Phi _{0}\xi _{c}^{-}}\left( \ln
\left( \frac{H_{c}\left( \xi _{ab}\right) ^{2}}{\Phi _{0}}\right) +c_{1}\right),
\label{eq12}
\end{equation}
according to Eqs. (\ref{eq3}), (\ref{eq10}) and (\ref{eq11}). Thus, given
the magnetization data of a homogenous system, attaining the limit $z=H_{c}\left( \xi _{ab0}^{\pm }\right) ^{2}\left\vert t\right\vert ^{-2\nu
}/\Phi _{0}<<1$, the growth of $\xi _{ab}$ and $\xi _{c}$ is unlimited and
estimates for $\xi _{c0}^{-}$ and $\xi _{ab0}^{-}$ can be deduced from
\begin{equation}
\left\vert t\right\vert ^{-2/3}\frac{m}{T}=-\frac{Q^{-}c_{0}^{-}k_{B}}{\Phi
_{0}\xi _{c0}^{-}}\left( \ln \left( \frac{H_{c}\left( \xi _{ab0}^{-}\right)
^{2}\left\vert t\right\vert ^{-4/3}}{\Phi _{0}}\right) +c_{1}\right) .
\label{eq13}
\end{equation}
In Fig. \ref{fig3} we depicted $\left\vert t\right\vert ^{-2/3}m/T$ \textit{vs}. ln$\left( \left\vert t\right\vert ^{-4/3}\right) $. From the straight lines we obtain
\begin{equation}
-\frac{Q^{-}c_{0}^{-}k_{B}}{\Phi _{0}\xi _{c0}^{-}}\simeq 0.025,
\label{eq14}
\end{equation}
and with that
\begin{equation}
\xi _{c0}^{-}\simeq 1.87\textrm{ \AA} .  \label{eq15}
\end{equation}
Furthermore, from $\ln \left( H_{c}\left( \xi _{ab0}^{-}\right) ^{2}/\Phi
_{0}\right) $ \textit{vs}. $\ln(H_{c})$ we deduce
\begin{equation}
\xi _{ab0}^{-}\simeq 15.6\textrm{ \AA .}  \label{eq16}
\end{equation}
For the anisotropy we obtain then the estimate
\begin{equation}
\gamma =\frac{\xi _{ab0}^{-}}{\xi _{c0}^{-}}\simeq 8.34,  \label{eq17}
\end{equation}
compared to $\gamma\simeq 13.4$, $\gamma\simeq 14.7$ \cite{kagawa} \ and $\gamma \simeq 12.3$ \cite{zech}.
\\
To explore the magnetic field induced 3D to 1D crossover further and to
probe the vortex melting line directly we invoke Maxwell's
relation
\begin{equation}
\left. \frac{\partial \left( C/T\right) }{\partial H_{c}}\right\vert
_{T}=\left. \frac{\partial ^{2}M}{\partial T^{2}}\right\vert _{H_{c}},
\label{eq18}
\end{equation}
uncovering the vortex melting transition in terms of a singularity, while the
magnetic field induced finite size effect leads to a dip. These features seem to
differ drastically from the nearly smooth behavior of the magnetization.
Together with the scaling form of the specific heat (Eq. (\ref{eq8})),
extended to the presence of a magnetic field,
\begin{equation}
c=\frac{A^{-}}{\alpha }\left\vert t\right\vert ^{-\alpha }f\left( x\right),
\textrm{ }x=\frac{t}{H_{c}^{1/2\nu }},  \label{eq19}
\end{equation}
we obtain the scaling form
\begin{eqnarray}
TH_{c}^{1+\alpha /2\nu }\frac{\partial \left( c/T\right) }{\partial H_{c}}&=&-\frac{k_{B}A^{-}}{2\alpha \nu }x^{1-\alpha }\frac{\partial f}{\partial x} \nonumber \\
&=&TH_{c}^{1+\alpha /2\nu }\frac{\partial ^{2}m}{\partial T^{2}}.  \label{eq20}
\end{eqnarray}
In Fig. \ref{fig4} we depicted $TH_{c}d^{2}m/dT^{2}$ \textit{vs}. $x$ for various
magnetic fields $H_{c}$. Apparently, the data collapses reasonably well on a
single curve. There is a peak and a dip marked by an arrow and a vertical
line respectively. Their occurrence differ clearly from the traditional mean-field
behavior where $\partial ^{2}m/\partial T^{2}=0$. The finite depth of the
dip is controlled by the magnetic field induced finite size effect. It
replaces the reputed singularity at the upper critical field obtained in the Gaussian
approximation \cite{prange}. Note that both, the peak and the dip are hardly
visible in the magnetization shown in Fig. 2. There we marked the location
of the peak, the dip and $T_c$ in terms of dashed and solid arrows.
The location of the dip determines the line
\begin{equation}
x_{p}=t_{p} H_{c}^{-3/4}\simeq -2.85\cdot 10^{-5}\textrm{ Oe}^{-3/4},  \label{eq21}
\end{equation}
in the $(H_{c},T)$-plane where the 3D to 1D crossover occurs. Along this
line, rewritten in the form $H_{pc}\left( T\right) =\Phi _{0}/\left( a\left(
\xi _{ab0}^{-}\right) ^{2}\right) \left( 1-T/T_{c}\right) ^{4/3}$, the
in-plane correlation length is limited by $L_{H_{c}}$(Eq. (\ref{eq1})). In
addition there is a peak at
\begin{equation}
x_{m}= t_{m} H_{c}^{-3/4}\simeq -8.35\cdot 10^{-5}\textrm{ Oe}^{-3/4}\textrm{,}
\label{eq22}
\end{equation}
corresponding to the vortex melting transition. Rewritten the vortex melting line follows in the form $H_{mc}\simeq2.7\cdot 10^{5}\textrm{ Oe}\cdot \left( 1-T_{m}/T_{c}\right) ^{4/3}$ which agrees very well
with the previous estimate $H_{mc}\simeq 1.8\cdot 10^{5}\textrm{ Oe}\cdot \left(
1-T_{m}/T_{c}\right) ^{4/3}$ obtained by Katayama \textit{et al}. \cite{katayama}, as the temperature dependence is concerned. Accordingly, we obtain the universal
ratios of the scaling variables of the reduced temperatures for the vortex melting
line and the 3D to 1D crossover line as
\begin{eqnarray}
\frac{z_{m}}{z_{p}}=\left( \frac{t_{p}\left( H_{c}\right) }{t_{m}\left(
H_{c}\right) }\right) ^{2\nu }&\simeq& 0.24,\nonumber\\
\textrm{ }t_{p}\left( H_{c}\right)
/t_{m}\left( H_{c}\right) &\simeq& 0.34.  \label{eq23}
\end{eqnarray}
These values agree well with the estimates $t_{p}\left( H_{c}\right)
/t_{m}\left( H_{c}\right) \simeq 0.3$ for a NdBa$_{2}$Cu$_{3}$O$_{7-\delta }$
single crystal \cite{plackowski} and $t_{p}\left( H_{c}\right) /t_{m}\left(
H_{c}\right) \simeq 0.35$ for a YBa$_{2}$Cu$_{3}$O$_{6.97}$ single crystal
\cite{roulin} derived from the respective references.
Finally, invoking the universal relation (\ref{eq9}) we obtain with $T_{c}=79.6$ K and $\xi _{c0}^{-}\simeq 1.87$ \AA\ (Eq. (\ref{eq15})) for the critical amplitude of the in-plane magnetic field penetration depth the value $\lambda _{ab0}\simeq 1.37\cdot 10^{-5}$ cm, in reasonable agreement with the estimate $\lambda _{ab0}\simeq 1.7\cdot 10^{-5}$ cm obtained from
magnetization data of polycrystalline YBa$_{2}$Cu$_{4}$O$_{8}$ samples \cite{khasanov}.\\

\section{\label{sec:level1}Summary}
We have shown that the analysis of reversible magnetization data of a YBa$_{2}$Cu$_{4}$O$_{8}$ single crystal provides considerable insight into the effect of thermal fluctuations and the magnetic field induced 3D to 1D crossover. In particular we demonstrated that the fluctuation dominated regime is
experimentally accessible and uncovers remarkable consistency with 3D-xy
critical behavior. Furthermore there is, however, the magnetic field induced finite size
effect. It implies that the correlation length transverse to the magnetic
field $H_{i}$, applied along the $i$-axis, cannot grow beyond the limiting
magnetic length $L_{H_{i}}=\left( \Phi _{0}/\left( aH_{i}\right) \right)
^{1/2}$, related to the average distance between vortex lines. Invoking the
scaling theory of critical phenomena clear evidence for this finite size
effect has been provided. In type II superconductors it comprises the 3D to
1D crossover line $H_{pi}\left( T\right) =\left( \Phi _{0}/\left( a\xi
_{j0}^{-}\xi _{k0}^{-}\right) \right) (1-T/T_{c})^{4/3}$ with $i\neq j\neq k$
and $\xi _{i0,j0,k0}^{-}$ denoting the critical amplitude of the correlation
length below $T_{c}$. As a result, below $T_{c}$ and above $H_{pi}\left(
T\right) $ superconductivity is confined to cylinders with diameter $L_{H_{i}}$(1D). Accordingly, there is no continuous phase transition in the $(H_c,T)$-plane along the $H_{c2}$-lines as predicted by the mean-field treatment. In addition, we confirmed the universal relationship between the 3D to 1D crossover and vortex melting line. The universal relation (\ref{eq9})
and Maxwell's relation (\ref{eq18}) also imply that the effects of isotope
exchange and pressure on $T_{c}$, in-plane magnetic field penetration depth,
correlation lengths, specific heat, and magnetization are not independent.

\section{Acknowledgments}
This work was supported by the Swiss National Science Foundation and in part by the NCCR program MaNEP.

\end{document}